\documentclass{article}
\usepackage{graphicx} 
\usepackage[margin=1in]{geometry}
\usepackage{url}
\usepackage{tcolorbox}
\usepackage{caption}
\usepackage{newfloat}
\usepackage{amsmath}
\usepackage[colorlinks=true]{hyperref}

\title{Generative AI and Security Operations Center Productivity: Evidence from Live Operations}
\author{James Bono, Justin Grana, and Alec Xu \\ \\ Microsoft Corporation }
\date{November 2024}

\begin{document}

\maketitle

\begin{abstract}
We measure the association between generative AI (GAI) tool adoption and security operations center productivity.  We find that GAI adoption is associated with a 30.13\% reduction in security incident mean time to resolution.  This result is robust to several modeling decisions.  While unobserved confounders inhibit causal identification, this result is among the first to use observational data from live operations to investigate the relationship between GAI adoption and security worker productivity.  
    \end{abstract}

\section{Introduction and Background}
Recent developments in generative artificial intelligence (GAI) have raised questions regarding its productivity effects. While GAI-based productivity enhancements have important labor market implications \cite{ailit1,ailit2}, 
its importance in cybersecurity extends beyond labor cost savings and has the potential to reduce costly attacks.  Despite its importance, there is scant evidence detailing the potential benefits of GAI tools for security professionals in live operations.

We present the first statistical evidence of productivity enhancements from GAI tools accruing to security operations center (SOCs) analysts in live operations.  Specifically, we conduct a difference-in-differences analysis of security incident mean time to resolution (MTTR) where the treatment is defined as adoption of Microsoft Security Copilot (``Copilot''), which ``combines a specialized language model with security-specific capabilities''\cite{cfs}.  Our results indicate Copilot is associated with a $30.13$\% reduction in MTTR three months post-adoption relative to a control group in the same period.  Although selection into treatment inhibits causal identification, our estimated effect on productivity aligns in magnitude with other recent work that leverages randomization to identify GAI's causal effects on knowledge-worker productivity \cite{defender1,git1,field1}.  Given that recent estimates suggest analysts spend, on average, 2.7 hours per day resolving incidents, costing \$3.3B in the US alone \cite{vect}, our results suggest that GAI tools have the potential to offer SOCs significant time and cost savings.  

\subsection{Related Work}
From a productivity perspective, our work sits adjacent to recent studies that examine the impact of GAI on worker productivity utilizing randomization in either a laboratory or field setting \cite{defender1,git1,field1,exp1,leg1}.  While lab experiments provide solid causal identification, whether they generalize to real world settings is a matter of judgment about the experimental design. Field experiments make a similar trade-off to lab studies. Random assignment gives them strong causal identification, but because they are often limited to a small set of organizations, the measured effects may not generalize more broadly. Our work makes the opposite trade-off; we examine telemetry data from live operations from over 150 organizations (see section \ref{sec:Data} below), so our results are likely to generalize to live operations for a broad range of organizations. However, our causal claims are weaker than lab and field experiments due to our inability to control for selection into treatment. Our work is more related to the limited studies that use observational data \cite{genai_live,git2} in live operations to estimate the impact of GAI on productivity.  As is typical in these studies, unconfoundedness cannot be guaranteed and thus our results provide evidence for ---  though do not causally identify --- GAI's impact on security worker productivity.  Despite their different shortcomings, studies have yielded surprisingly consistent estimates of GAI's productivity effects across domains (see table \ref{tab:sum} for specific estimates) with our 30.13\% estimate being similarly consistent.

\begin{table}[h]
\centering
\begin{tabular}{|p{1cm}|p{7cm}|p{7cm}|} \hline
\textbf{Source} & \textbf{Domain} & \textbf{Estimated Productivity Gain}  \\ \hline 
\cite{genai_live} & Customer technical support & 34\% reduction in task completion time for novices\\ \hline 
\cite{defender1} & Security incident laboratory experiment & 23\% decrease in task completion time \\\hline 
\cite{git1} & Laboratory experiment implementing HTTP server in Javascript & 55.8\%  decrease in time to completion\\ \hline 
\cite{field1} & Field study of software development tasks & 26.08\% increase in tasks completed \\ \hline
\cite{itrct} & Laboratory study of IT Admins & 34.53\% accuracy improvement and 30.69\% reduction in task completion time \\ \hline
\end{tabular}
\caption{Estimated Productivity Gains from Generative AI}
\label{tab:sum}
\end{table}

From the security perspective, our study builds on research about automation of security tasks. The literature on this topic finds that risk mitigation is essential with the growth of security breaches, especially because so many security vulnerabilities operate at the gap between how systems are supposed to operate and how they actually operate \cite{morgan}. Although many aspects of cybersecurity currently rely on human subject matter experts \cite{costa}, researchers point out the possibility of automating error-prone and time-consuming security work. Machine learning techniques show particular promise in intelligently analyzing cybersecurity data \cite{sarker}. Natural language processing, knowledge representation and reasoning, and rule-based expert systems modeling can also support AI-driven cybersecurity \cite{sh}.  

\subsection{Security Operations Center and Copilot}

\paragraph{What is Security Event Management?} Security event triage and response is a a central function of an organization's cybersecurity operations \cite{event}.  Broadly, computer network activity generates telemetry (logs, signatures, etc).  That telemetry is then collated and processed by tools such as security information and event management (SIEM) and extended detection and response (XDR) solutions. These solutions often generate security alerts based on logical rules or statistics and machine learning models.  Ultimately, SIEM and XDR solutions aggregate telemetry into a set of distinct units, representing holistic depictions of suspicious activity, for human analysts to investigate.  We adopt the convention of referring to these aggregate units as ``security incidents'' or just ``incidents.'' 

Teams of analysts triage and resolve incidents.  Some incidents may be resolved as a ``false positive'' where the suspicious behavior is determined to be benign.  Other incidents may require more intensive remediation and response such as altering users' privileges, disconnecting compromised systems, removing malicious files, and applying patches. Once an analyst resolves an incident, they can proceed to the next incident in their SIEM or XDR solution. However, as noted above, the incident arrival rate far surpasses what a typical SOC can effectively triage. Recent estimates suggest that as much of 67\% of security incidents go unresolved \cite{vect}.  

\begin{figure}[ht!]
    \centering
    \includegraphics[width=0.9\linewidth]{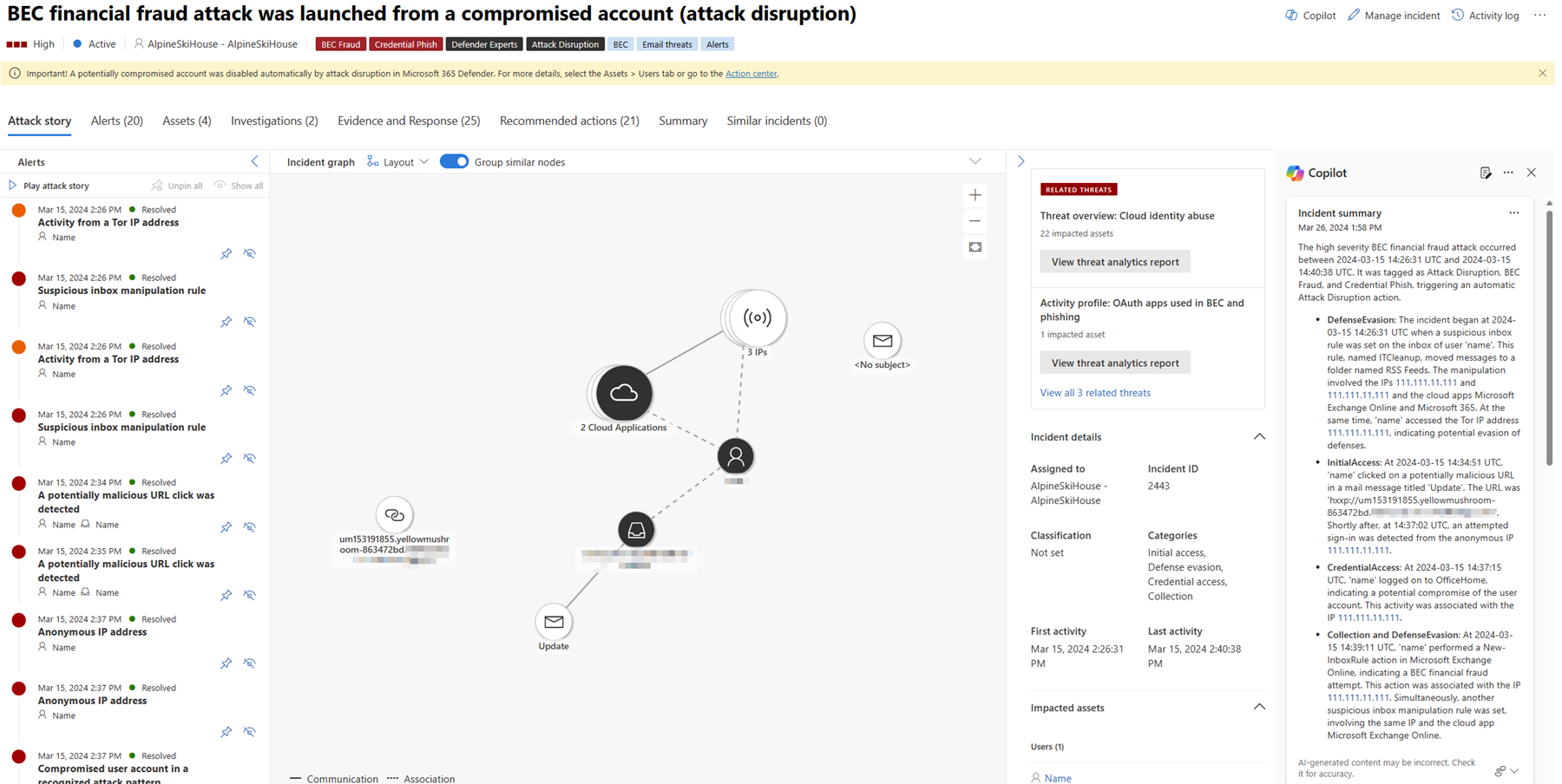}
    \caption{ An example of incident summarization for a Business Email Compromise Incident}
    \label{fig:sc_ex}
\end{figure}

\paragraph{What is Copilot?} Microsoft Copilot is a GAI tool designed to enhance the operational efficiency and effectiveness of SOCs.  One salient feature related to incident resolution is incident summary.  Previously, analysts would need to investigate each piece of information independently to obtain the full breadth of an incident.  Copilot instead aggregates that information and serves it to the user in human-readable format.  The right panel in Figure \ref{fig:sc_ex} illustrates an example of alert summarization for a Business Email compromise incident. 

Although a full list of the product features is beyond the scope of this paper, we mention the following additional features to give the reader a better understanding of the mechanisms by which Copilot affects MTTR:

\begin{itemize}
    \item Guiding analysts to the appropriate response actions based on the incident data
    \item Helping interpret malicious scripts associated with the incident
    \item Creating scripts to query security logs based on natural language inputs
    \item Retrieving relevant threat intelligence using natural language inputs
\end{itemize}

\section{Data and Method}
\label{sec:Data}
We leverage system metadata generated by the Microsoft Defender XDR product spanning the 180 days beginning March 3, 2024 and ending August 30, 2024.
 To get incident time to resolve, we compute the difference between the time an incident is first opened by an analyst and the time the incident is marked as resolved. We also leverage the following incident metadata:

\begin{tcolorbox}[title=Box 1: Incident Metadata]
\begin{itemize}
\item Whether an incident is high severity
\item The number of alerts aggregated to define the incident 
\item Microsoft Security products that triggered alerts.  They can be
\begin{itemize}
\item Microsoft Defender for Endpoint
\item Microsoft Defender for Identity
\item Microsoft Defender for Cloud Apps
\item Microsoft Defender for Office
\item Data Loss and Protection
\item Sentinel
\item App Governance
\item Entra ID
\item Defender XDR 
\end{itemize}
\end{itemize}
\end{tcolorbox}

Organizations that have consumed Copilot resources are considered treated. To determine treatment times, we use the date the organization first consumed Copilot resources. For each organization we also include the country and industry as organization-level metadata. Finally, we use the number of licenses of the above Microsoft Security products an organization has to conduct our propensity score matching (described below).

\subsection{Method}

Our approach proceeds in two steps.  First, after identifying our treatment group as those that consume Copilot resources, we use propensity score matching to construct a control group.  To do this, we conduct a binary logistic regression of Copilot adoption on organization segment, industry, country, strategic category, and available seats per Microsoft security product. We then match each organization that has adopted Copilot with an organization that has the closest estimated probability of adoption.  

In the second step, we estimate a difference-in-differences model with the log of incident resolution times as the independent variable.  We only consider incidents that are resolved within 12 hours after opening. We limit the time to resolve because longer incidents can be driven by a variety of factors that are beyond what we expect our treatment variable, Copilot, to affect, e.g., shift changes and new priorities. We also limit our treatment group to organizations that have conducted at least 200 Copilot executions and have at least 15 observations in the periods before and after treatment. Nevertheless, we show our results are robust to these key modeling decisions.  After filtering, our analysis considers 89 Copilot adopters in the treatment group and 88 non-adopters (two adopters matched to the same non-adopter). There was a total of 95,522 incidents with 52,698 of those observations for the treatment group and 42,824 of those observations for the control group.  

The ``before'' period includes incidents that were closed before the organization first uses Copilot and opened at least 12 hours before their first Copilot use. The ``after'' period includes incidents that started between the time of the first Copilot use and 12 hours before the end of the data. The before and after periods for the control group are determined by the before and after periods of their treatment group counterparts. We exclude incidents that start between 12 hours pre-adoption and adoption as well as incidents 12 hours before the end of data to control for effects due to censoring as well as incidents that overlap the treatment periods.  

The difference-in-differences regression specification is standard with two exceptions.  First, we include weekly fixed effects, organization industry and country, and the incident metadata given in Box 1.  
Second, we disaggregate the after period by month to discern the relationship between Copilot adoption and MTTR one, two, and three or more months after adoption.  

Following \cite{Angrist2009,Ciani2020}, we model the logarithm of time to resolve because a multiplicative common trends assumption makes more sense than an additive one. That is, common trends should reflect the relative change in difficulty to resolve an incident due to the evolving threat landscape. For example, take two organizations, the first with an MTTR of 10 and the second with an MTTR of 100, both in the before period. Then, if the first has an MTTR of 12 in the after period, multiplicative common trends says the second would have an MTTR of 120 in the after period, not 102, as additive common trends would imply. Time to resolve also roughly follows a lognormal distribution, so the log specification offers a much better model fit the alternative specification in levels. 

Formally, we estimate the following model:

\begin{equation}
\log(TTR_{it})= \alpha + \eta_1I[Treatment_i] + \sum_{\tau=1}^3\big[\gamma_tI[\tau=t] + \beta_tI[\tau=t]\times I[Treatment_i]\big] + \text{controls}
\end{equation}
where 
\begin{itemize}
\item $i$ indexes incidents
\item $t$ indexes months after adoption
\item $I[Treatment_i]$ indicates if the incident originated from an organization that adopted Copilot.
\end{itemize}
The $\beta_t$ parameters capture the difference-in-differences between the MTTR of the treatment and control group $t$ months after adoption.  When reporting results, we present  $1-e^{\beta_t}$ which approximates the percentage reduction of MTTR associated with treatment. See appendix \ref{sec:ap1} for details of this approximation. This model is what \cite{dyn_did} refers to as the ``dynamic specification'' of a two-way fixed effects model but removes the organization level fixed effects.  Section \ref{sec:robustness} discusses extending the model to include such fixed effects.

\section{Results}

Our main result is that Copilot adopters experienced a statistically significant 30.13\% reduction in mean time to resolution three months post adoption relative to the control group.  The results are summarized in table \ref{tab:restab}.  

\begin{table}[h!]
\centering
\begin{tabular}{|p{3cm}|p{3cm}|p{3cm}|p{3cm}|} \hline
\textbf{Quantity} & \textbf{Estimate} & \textbf{S.E. (clustered)} & \textbf{p-value}  \\ \hline 
$1-e^{\beta_1}$ & -.0301 & .1295 & .8191 \\ \hline
$1-e^{\beta_2}$ & .1293 & .1427 & .3956 \\ \hline
$1-e^{\beta_1}$ & .3013 & .1806 & .0487 \\ \hline
\end{tabular}
\caption{Reduction in MTTR for treatment group.  Positive numbers represent a reduction.}
\label{tab:restab}
\end{table}

\begin{figure}[h!]
    \centering
    \includegraphics[width=0.6\linewidth]{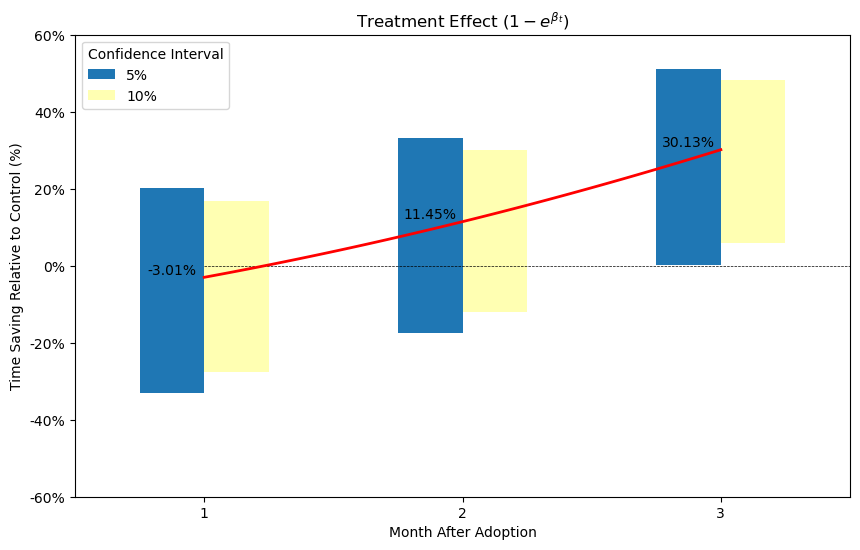}
    \caption{ Parameter estimates for treatment effect one, two, and three months after treatment.}
    \label{fig:res1}
\end{figure}

Although we do not observe statistically significant differences in the first and second months post adoption, the estimated coefficients trend toward increasing gains over the three-month period. The delay in effect suggests that organizations need time to learn about how to best integrate Copilot in their workflows. However, this analysis does not rule out other reasons for the delay such as the time it takes for analysts to adopt Copilot across an organization. The estimates for all three months and their 95\% (blue) and 90\% (yellow) confidence intervals are given in figure \ref{fig:res1}

\subsection{Robustness}
\label{sec:robustness}
In general, our results are robust at the 10\% significance level for many alternative specifications. As Table \ref{tab:robust} shows, the results are largely robust at the 10\% significance level with estimates ranging from 22.60\% to 31.84\%. The only specification that loses statistical significance is when we remove the minimum incident time. However, even there, we easily regain statistical significance by restricting the dataset to organizations that have 20 (compared to 15) observations before and after treatment (effectively removing organizations that provide little signal due to lack of observations), with an estimated effect of 30.20\% and a p-value of .0625.  Furthermore, if we limit the study to organizations that have closed at least 100 incidents in the before and after period and perform a two-way fixed effects regression with monthly fixed effects and  standard errors clustered\footnote{Recent work \cite{cluster} calls into question whether standard errors should be clustered at all. 
 However, we cluster standard errors to be conservative and all of our results remain statistically significant if we do not cluster standard errors.} at the organization-time level, we get a similar estimated time savings of 33.69\% (p=.039).

\begin{table}[h!]
\resizebox{\textwidth}{!}{%
\begin{tabular}{|l|ll|ll|}
\hline
                                             & \multicolumn{2}{c|}{Weekly Fixed Effects}            & \multicolumn{2}{l|}{Monthly Fixed Effects}           \\ \hline
Specification                                & \multicolumn{1}{l|}{Relative Time Savings} & p-value & \multicolumn{1}{l|}{Relative Time Savings} & p-value \\ \hline
Additionally control for Alert Categories in Incidents.                               & \multicolumn{1}{l|}{28.57\%} & .061* & \multicolumn{1}{l|}{29.28\%} & .0558* \\ \hline
Remove incident minimum time                 & \multicolumn{1}{l|}{22.60\%}               & .164    & \multicolumn{1}{l|}{23.71\%}               & .1427   \\ \hline
\begin{tabular}[c]{@{}l@{}}Increase Usage Threshold \\ To 400 Executions\end{tabular} & \multicolumn{1}{l|}{30.23\%} & .055* & \multicolumn{1}{l|}{31.93\%} & .051*  \\ \hline
Reduce Copilot Usage Threshold to 100 Executions & \multicolumn{1}{l|}{30.44\%}               & .046**  & \multicolumn{1}{l|}{31.15\%}               & .040**  \\ \hline
Increase Incident Maximum time to two days   & \multicolumn{1}{l|}{31.84\%}               & .064*   & \multicolumn{1}{l|}{33.11\%}               & .054*   \\ \hline
\end{tabular}%
}
\caption{Estimate of time savings associated with the treatment group three months after adoption ($1-\beta_3$) for alternative modeling specifications. ** indicates significance at the 5\% level where * indicates significance at the 10\% level.}
\label{tab:robust}
\end{table}

\section{Discussion}
Our main result is a 30.13\% reduction in MTTR associated with Copilot adoption.  This 30.13\% is similar to other studies that causally estimate productivity enhancements stemming from GAI adoption.  Our statistical model is robust to alternative modeling decisions. 

Although we use propensity score matching to control for selection into treatment based on observables, it is impossible to rule out the possibility that factors other than Copilot are generating the increase in productivity. For example, an organization may adopt Copilot because the SOC received a budget expansion. So, even though Copilot may contribute to productivity gains, the co-occurrence of unobserved factors like an increased number of analysts, other new software licenses, and additional trainings make it impossible to isolate the impact of Copilot on productivity using observed telemetry alone.

Moreover, it is even likely that those that are willing to pay for Copilot are the ones that would most benefit from Copilot. In other words, when organizations perform a cost/benefit analysis on the decision to adopt Copilot, only those with the highest benefit would adopt Copilot, since all organizations face the same sticker price. This means that, even assuming our estimates are causal, our estimates would overstate the average productivity impact of Copilot for \textit{non-adopters}.

Considering these limitations, our results provide evidence for the causal impact of Copilot on productivity as measured by MTTR. However, short of performing a randomized control trial on live organization operations or discovering a natural experiment that mimics randomized selection, the associations uncovered in this study are among the most promising signals supporting the positive effect of GAI tools (like Copilot) on the productivity of security operations. Furthermore, the results are statistically significant and robust, so it is clear that those that adopted Copilot observed a productivity gain.

Future research can help control for some of the unobserved confounders.  Specifically, continuing to move from a basic difference-in-differences specification to a two-way fixed effects model can help address selection issues by leveraging heterogeneous adoption times and organization fixed effects.  The current volume of available data does not provide sufficient power for a two-way fixed effects model since many organizations only have a small number of incidents.  Despite this, the fact that our point estimate when conducting a two-way fixed effects regression on large organizations shows a similar magnitude of effect as our main result  is an encouraging signal.  

\bibliographystyle{plain}
\bibliography{refs}

\appendix
\section{Appendix}
\label{sec:ap1}

Consider a standard two period, two group difference-in-differences regression where $y_{td}$ is the response variable indexed by time period ($t$) and whether an individual was treated ($d$).  Let $\beta$ be the parameter on the $Treatment\times Time$ interaction variable.   As \cite{logs} notes, in a log-linear difference in difference regression 
\begin{equation}
    e^{\beta} = \frac{\frac{y_{11}}{y_{01}}}{\frac{y_{10}}{y_{00}}}
    \end{equation}
    However, when the response of the treatment and control group are of similar magnitudes in the before period so that $y_{00} \approx y_{01}$, we get
    \begin{equation}
        1-e^{\beta} \approx \frac{y_{11}-y_{10}}{y_{01}}
    \end{equation}
which represents the percent reduction in the response associated with treatment.  Furthermore, if the control group does not change over time ($ y_{10} \approx y_{00}$), then
 \begin{equation}
        1-e^{\beta} \approx \frac{y_{11}-y_{01}}{y_{01}}
    \end{equation}
    which again represents the percent reduction associated with treatment.  

    In our data, the treatment and control group have a similar MTTR before treatment (58 minutes for treated, 53 minutes for control) and the control group's MTTR changes minimally in the third month in the post treatment period  (52 minutes). Hence, we use $1-e^\beta$ as an approximation for percent reduction in MTTR due to treatment.  

\end{document}